\documentclass[12pt]{JHEP3}

\title{Pentaquark hadrons from lattice QCD}

\author{F.~Csikor$^a$, Z.~Fodor$^{a,b}$, 
        S.D.~Katz$^c$\footnote{On leave from Institute for Theoretical 
          Physics, E\"otv\"os University, Hungary}  
    and T.G.~Kov\'acs$^b$\footnote{On leave from Department of Theoretical 
        Physics, University of P\'ecs, Hungary}  \\
    $^a$Institute for Theoretical Physics, E\"otv\"os University, Hungary \\
    $^b$Department of Physics, University of Wuppertal, Germany\\
    $^c$Deutsches Elektronen-Synchrotron, DESY Hamburg, Germany }

\abstract{We study spin 1/2 isoscalar and isovector candidates in both
parity channels for the recently discovered
\Tp(1540) pentaquark particle in quenched lattice QCD. 
Our analysis takes into account all possible 
uncertainties, such as statistical, finite size and quenching
errors when performing the chiral and continuum extrapolations and
we have indications that our signal is separated from scattering states.
The lowest mass that we find in the $I^P=0^-$ channel is 
in complete agreement with the experimental value of the \Tp\ mass.
On the other hand, the lowest mass state in the opposite 
parity $I^P=0^+$ channel is much higher. Our findings
suggests that the parity of the \Tp\ is negative.}

\preprint{DESY-03-140 \\ ITP-Budapest-600 \\ WUB-03-09}

\newcommand{\msub}[1]{\ensuremath _{\mbox{\scriptsize #1}}} 
\newcommand{\Tp}{\ensuremath $\Theta^+$}
\newcommand{\be}{\begin{equation}}
\newcommand{\ee}{\end{equation}}

\usepackage{graphics}

\begin{document}

\section{Introduction}

The possible existence of exotic hadrons has long been put forward in
several contexts, but the experimental discovery
of the first such particle had to wait until early
this year \cite{Nakano:bh}-\cite{Barmin:2003vv}. For a long time exotic hadron
states containing more than three quarks were not considered 
to be of great importance on account of their presumed large
decay widths. (For a recent discussion of this issue see e.g.\ 
\cite{Gothe:2003sw}.) The experimental discovery of the 
$\Theta^+(1540)$ particle changed this situation dramatically.
Indeed, the tightest experimental upper bound so far 
on the width of the $\Theta^+$
is around 10~MeV \cite{Barmin:2003vv}, and based on the reanalysis of older
experimental data, more stringent constraints on the width are 
also suggested \cite{width}. This remarkably narrow width 
would explain why the $\Theta^+$ has not been seen before.

Since the \Tp\ decays into a neutron and a $K^+$, its strangeness has to be  
+1, the third component of its isospin is 0, and its minimal quark content 
is $dduu\bar{s}$. From the lack of a signal in the $I_3\!=\!1$ channel
the SAPHIR collaboration concluded that the \Tp\ should be an isospin
singlet state \cite{Barth:2003es}. Its spin and parity cannot be 
unambiguously pinned down based on currently available experimental data. 

The experimental discovery of the \Tp\ pentaquark triggered a flurry of 
theoretical speculations \cite{livelydebate} 
about its possible structure, yet unmeasured 
quantum numbers and on the possibility of the existence of other 
exotic hadrons. Chiral soliton models 
\cite{soliton}, as well as correlated quark models 
\cite{Jaffe:2003sg,diquark,quarkmodel}
suggest that the \Tp\ should be a spin 1/2 positive parity
isosinglet state.  In contrast, uncorrelated quark 
models predict negative parity. QCD sum rules have also 
been used to study the \Tp\ \cite{sumrules}. 

In the present work we use lattice QCD to study the properties of
some pentaquark states that are likely candidates to be identified
with the \Tp. This is the only available tool to 
extract low energy hadronic properties starting from first 
principles, i.e.\ QCD. In our exploratory study we choose to work
in the quenched approximation, which is known to be successful
in reproducing mass ratios of stable hadrons
\cite{Gattringer:2003qx}-\cite{Shanahan:1996pk}.

Working at three different values of the lattice spacing,
$a=0.17,0.12$ and $0.09$~fm enables us to do a continuum extrapolation.
Since using light enough quarks that reproduce the physical
pion mass would be prohibitively costly, for each lattice
spacing we simulate at four different, somewhat larger 
pion masses and extrapolate to the physical limit. 
This is done in two different ways, by
keeping the $s$-quark mass fixed to reproduce the physical
kaon mass in the chiral limit and also by changing the $s$-quark
mass in a way to keep the nucleon to kaon mass ratio fixed as 
the light quark mass is varied.  In order to have  finite size effects under
control we always keep the linear physical size of the spatial box fixed
at  $L\approx 2$~fm. In addition, on the two coarsest lattices, 
we also perform a finite volume analysis by considering different 
volumes.

Since the exact structure of the \Tp\
is yet unknown, there are quite a few different possibilities
to construct a baryon with the given quark content $dduu\bar{s}$.
In our computation we use two particular spin 1/2 
operators, one with isospin $I=0$ and and another with $I=1$. 
The analysis is carried out in both parity
channels.

The main result of the present work is that 
the lowest mass that we find in the $I^P=0^-$ channel is
in complete agreement with the experimental value of the \Tp\ mass.
On the other hand, the lowest mass in the opposite
parity $I^P=0^+$ channel turns out to be much larger, thus
we suggest negative parity for the \Tp\ particle.

The lightest state in the $I=1$ channel also has negative
parity and comes out to be 
about 15\% heavier, but still within one $\sigma$ of the \Tp\ mass.
The SAPHIR Collaboration, however, have 
already ruled out the \Tp\ to be a member
of an $I\!>\!0$ multiplet \cite{Barth:2003es}, 
therefore the triplet state that we 
found might be a genuinely new state not seen before, albeit with 
a potentially much larger width than that of the \Tp.

The rest of the paper is organized as 
follows. In Section \ref{se:sd} we give the details
of our simulations and our data analysis. 
In Section \ref{se:c} we present our results and conclusions.
Those who are not interested in lattice technicalities 
can directly skip to Section \ref{se:c}.

\section{Simulation details and data analysis}
     \label{se:sd}

In this section we discuss some technical details of our lattice 
simulations. We use the Wilson gauge action and the Wilson 
fermion action. The only ``non-standard'' piece of the calculation
is the choice of the pentaquark operator.  Since there is general 
agreement in the literature that the \Tp\ is very likely to have
spin 1/2, we restricted ourselves to this choice. Nevertheless,
it will be interesting to look at higher spin states too. After fixing the
quark content of the particle to be $dduu\bar{s}$ and its spin
to 1/2, there is still considerable freedom 
in choosing the interpolating operator. Even though the \Tp\ is 
experimentally suggested to be an isosinglet, we computed correlators both 
in the $I=0$ and $I=1$ channels using the interpolating fields
\be
  \eta_{0/1} = \epsilon^{abc}[u_a^T C\gamma_5 d_b] 
           \{u_e \bar{s}_ei\gamma_5 d_c \mp (u \leftrightarrow d) \}.
     \label{eq:eta}
\ee
Here $C$ is the charge conjugation
operator, only colour indices are explicitly written out and the 
$I=0$ case corresponds to the minus sign. There are many more 
possibilities even in the I=0,1 channels. 
In principle testing other interpolating 
operators and checking which one has the best overlap with
the \Tp\ state should provide information on the wave function
of the particle. In this exploratory study, however, we do not pursue this 
direction any further. 

We did not study the $I=2$ channel either,
mainly because it seems to be computationally more expensive.
At first sight this is hard to understand since
normally in lattice simulations one is used to quark propagator
computations being the most expensive part. Given the propagators,
calculating the different contractions entering hadron correlators
takes only a tiny fraction of the time. With these more complicated
pentaquark hadrons, however, that is no longer the case. Already in the 
$I=0,1$ channels the time for computing the contractions is of the same 
order of magnitude as the fermion matrix inversion. The reason for that
is mainly that the number of terms occurring in the colour and Dirac
index sums grows exponentially with the number of constituent particles.

\TABLE{
%\begin{center}
%\vspace{3mm}
\begin{tabular}{|l||r|r|r|r|} \hline
 $\beta$ & size              & $a$ (fm) & L (fm) & confs  
                                                            \\ \hline\hline
  5.70   & $10^3\times 24$   &  0.171   & 1.71   & 200      \\ \hline
  5.70   & $12^3\times 24$   &  0.171   & 2.05   & 220      \\ \hline
  5.70   & $16^3\times 32$   &  0.171   & 2.73   & 100      \\ \hline
  5.85   & $10^3\times 32$   &  0.123   & 1.23   & 726      \\ \hline
  5.85   & $12^3\times 32$   &  0.123   & 1.48   & 348      \\ \hline
  5.85   & $16^3\times 32$   &  0.123   & 1.97   & 90       \\ \hline
  6.00   & $20^3\times 36$   &  0.093   & 1.86   & 94       \\ \hline
\end{tabular}
%\end{center}
  \caption{The seven gauge field ensembles used for the study of the
pentaquark. The lattice spacing has been set with the Sommer parameter 
$r_0=0.50$~fm. 
      \label{tab:pars}}
}

The \Tp\ can decay into a kaon and a
nucleon even in the quenched approximation. The experimentally
very narrow pentaquark state is just slightly above the $N+K$ 
threshold, thus one expects that the \Tp\ is visible in correlators
of our well localised operators. Note that in principle scattering
states can mix in the spectrum, though they have a characteristic 
volume dependence. (This volume dependence can be used to calculate 
decay rates c.f. \cite{Lellouch:2000pv}). 
On the two coarsest lattices we searched for 
any possible volume dependence
of the \Tp\ by considering different spatial lattice sizes ranging
from $L=1.3$ to $2.7$~fm. We did not see the expected volume dependence 
characteristic to the S-wave scattering state. (Note, however that
this signature would appear only in the $I=1$ channel, since in
the $I=0$ channel the scattering length is too small.) 
In addition, on our largest lattice at the largest quark mass, based on 300
configurations, we computed the $2 \times 2$ 
correlation matrix of the $\eta$ operator
of Eq.\ (\ref{eq:eta}) and a similar operator with its colour indices 
contracted as in the nucleon and the kaon. The latter is expected to
couple more strongly to the scattering state. After analysing the 
cross correlator matrix
we obtained for the smallest energy state $E_0/(m_N+m_K)=0.994(18)$ (which
we interpret as the S-wave scattering state) and for the state above 
$E_1/(m_N+m_K)=1.074(20)$ (which is the \Tp\ candidate).
Note that the volume of our spatial box implies that the next
scattering state with zero total momentum should not occur below
$E/(m_N+m_K)=1.21$. Thus, in particular, we do not expect to be faced 
with the unlucky scenario that in our analysis one of the scattering 
states is accidentally lifted to the \Tp\ mass.
  
Given our choice of interpolating operators for the pentaquark, we 
now discuss other details of the simulations. We generated seven different
ensembles of quenched gauge field configurations 
with the Wilson plaquette action.
Quark propagators were calculated using Wilson fermions.
The parameters of our gauge ensembles are summarized in Table \ref{tab:pars}.
Ensembles with three different values of the coupling were 
chosen to have roughly the same spatial volume in order to avoid
any finite size contamination. 

The last ingredients of our computation are the choice of quark masses 
and the way we fix the two independent parameters, the lattice spacing and the
$s$-quark mass. Due to the extensive literature
on quenched spectroscopy one can safely plan the necessary
parameter sets in order to carry out acceptable chiral
and continuum extrapolations. We used light quark masses in the
range $m_\pi/m_\rho=0.5-0.7$. In quenched QCD there is a small inherent 
ambiguity in how those parameters can be fixed. Given the 
fact that the \Tp\ is intimately connected 
to the kaon, it seems reasonable to use the physical mass of the kaon
to set the $s$-quark mass. We adopted two different 
schemes in approaching the chiral limit.
In both cases the light quarks were taken degenerate. In scheme (A)
we kept the $s$-quark mass fixed to reproduce the physical kaon
mass in the chiral limit. In scheme (B) we changed the mass of the 
$s$-quark along with that of the light quarks so as to keep the
nucleon to kaon mass ratio constant. 

One of the main questions we set out to address in the present study is
to determine the parity of the \Tp\ particle. Simple baryon interpolating
operators in general are expected to couple to states of both parities
\cite{Montvay:cy,Sasaki:2001nf}. With the parity assignment of our
operators the correlators are of the form\footnote{In the first version
of this paper the signs of the terms proportional to $\gamma_0$ were
opposite. We thank S.~Sasaki \cite{Sasaki} and K.F.~Liu \cite{Liu}
for pointing out this apparent error. The only consequence of 
correcting this error is the flip of the parity assignment. 
All the other results, including correlators and masses remained unchanged.} 
\be
  \langle B_0 \bar{B}_t \rangle = 
  (1-\gamma_0) \left[ C_+ \mbox{e}^{-tm_+} + C_- b \mbox{e}^{-(T-t)m_-}\right]
 +(1+\gamma_0) \left[ C_+ b \mbox{e}^{-(T-t)m_+} + C_- \mbox{e}^{-tm_-}\right],
     \label{eq:corr}
\ee
where $T$ is the size of the lattice in the time direction, $b=\pm 1$
corresponds to periodic/antiperiodic boundary conditions in the time
direction, and $m_{\pm}$ are the masses of the $\pm$ parity states. 
To be able to project out the two parities separately, we
always computed both components of the correlators, i.e.\ the one proportional
to $1$ and $\gamma_0$. In addition, at one quark mass on each
ensemble we also computed the correlators with antiperiodic
boundary condition, which enabled us to determine the two
parity partners in another  way.

We performed two-parameter correlated fits to the $1$ and 
$\gamma_0$ components of the $\langle B_0 \bar{B}_t \rangle$
pentaquark correlators with the standard cosh and sinh form respectively,
assuming that at large enough separation the lightest state dominates the
correlators. Since our data sample was rather limited we followed 
Ref.\ \cite{Michael:1994sz} and
used correlated fitting with smeared smallest eigenvalues of the 
correlation matrix. We could always find fit ranges with acceptable
$\chi^2\approx 1$ accompanied by a plateau in the effective 
and in the fitted masses.
The fits to the $1$ and $\gamma_0$ components always 
gave compatible results. The interpolating operators, however, 
could be better chosen, since the mass plateaus were reached at 
relatively large time separations compared to e.g.\ the nucleon correlator. 
In Fig.\ \ref{fig:massfit} we show a typical set of fits in 
the $I=0$ channel. 

\FIGURE{
\resizebox{10cm}{!}{\includegraphics{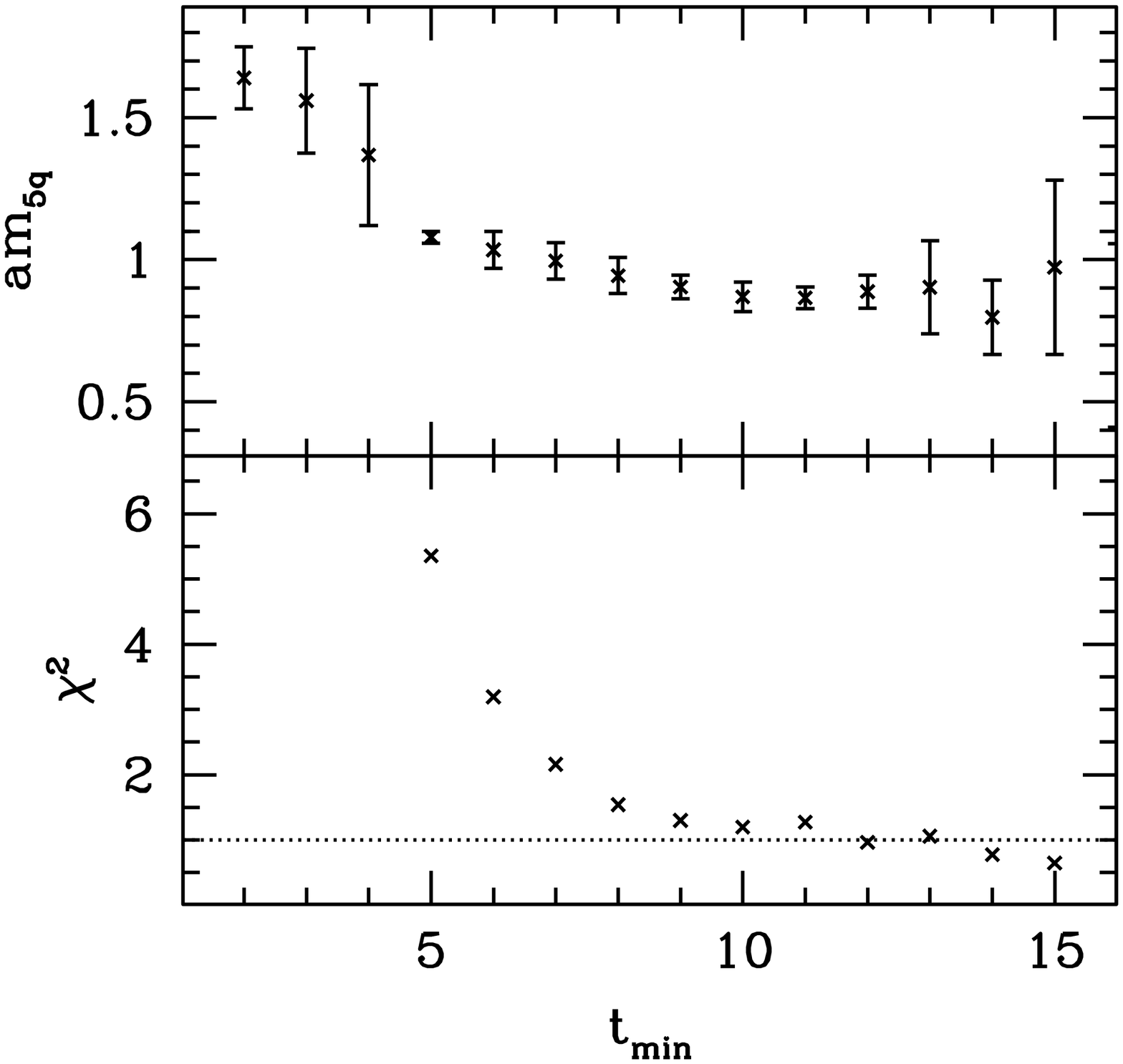}
\caption{\label{fig:massfit} A typical fit for a correlator in the 
$I=0$ channel at $\beta=5.85$ at our lightest $u$ quark. 
The mass given by the fit along with the corresponding
$\chi^2$ per degree of freedom is shown versus the starting point 
of the fit range.}}
}

Since one of our most important goals is to determine the parity,
let us discuss this issue in some detail.
The interpolating operators we used are expected to couple to states of
both parities in the given channel. In order to determine the parity
of the lowest state as well as the mass of its heavier parity partner
we performed single exponential fits to both the 
$1+\gamma_0$ and the $1-\gamma_0$ components of the correlators.
As can be seen from Eq.\ (\ref{eq:corr}) these projections should fall off
with two different exponents for $0 \leq t \ll T/2$ and for
$T/2 \ll t < T$, corresponding to the two different parity
states. As expected, we could always confirm that one of the exponents was
compatible with our previously found lowest mass obtained by 
the cosh and sinh fits to the $1$ and $\gamma_0$ components, respectively. 
Both for the $I=0$ and the $I=1$ channels, the negative parity state  
turned out to be the lighter one of the two parities.

The above correlators were calculated with periodic quark boundary 
conditions. An independent method to determine the parity
is based on the combination of the correlation functions obtained
with periodic and antiperiodic boundary conditions. We carried
out this analysis at each lattice spacing for the largest 
quark masses.  As can be
seen from Eq.\ (\ref{eq:corr}), the sum (difference) of the 1 component
of the periodic and the $\gamma_0$ component of the antiperiodic correlator
projects onto the negative (positive) parity state, respectively. Single
mass fits to these combinations yielded masses in complete agreement 
with our mass and parity determinations using periodic boundary
conditions.

A consistent 
picture emerged for the parity of the pentaquark states 
at all lattice spacings and quark masses.  
The negative parity mass was always by many standard deviations below
the positive parity mass. This result was obtained from both
boundary conditions combining the 1 and $\gamma_0$ channels
and also from combining different boundary conditions. 
Thus, we conclude that both for 
the $I=0$ and the $I=1$ channels, the negative parity state  
was determined to be the lighter one of the two parities.

The masses in the $I^P=0^\pm,1^\pm$ channels were then 
extrapolated to the chiral limit. As we already mentioned, for the chiral
extrapolations we used two different sets of quark masses and correspondingly
two different methods.
\begin{enumerate}
\item[(A)] We set the scale by the Sommer parameter $r_0=0.5$ fm and we 
         held the $s$ quark mass fixed to reproduce the physical kaon
         mass $m_K=494$ MeV. This approach resulted in
     strange quark hopping parameters of $\kappa_s$=0.1621,0.1574 and 0.1544
         for $\beta$=5.70,5.85 and 6.00, respectively.
         At all three lattice spacings four different light quark masses
         were used in order to extrapolate into the chiral limit.
\item[(B)] A somewhat different method to approach the chiral and continuum
         limit can be imagined as follows. For all the three lattice spacings
         we used four sets of strange and light hopping parameters. 
         The light quark masses were used to extrapolate into the
         chiral limit, whereas 
         we chose the strange quark mass to
         reproduce the experimental nucleon to kaon mass ratio.
\end{enumerate}
We followed both methods all the way through, first performing
the chiral extrapolation at fixed $\beta$ then carrying out the 
continuum extrapolation. The two methods (A and B) 
should give identical final results, though the difficulties
connected with the chiral and continuum extrapolations might be
quite different. Luckily enough, the extrapolations for the 
relevant mass ratios were equally easy for both techniques. The
obtained results are in complete agreement. Since we used
the same gauge configuration ensembles for the two methods, the 
results are strongly correlated and cannot be combined.
In what follows, we focus on the somewhat more 
conventional method (A).

\FIGURE{
\resizebox{10cm}{!}{\includegraphics{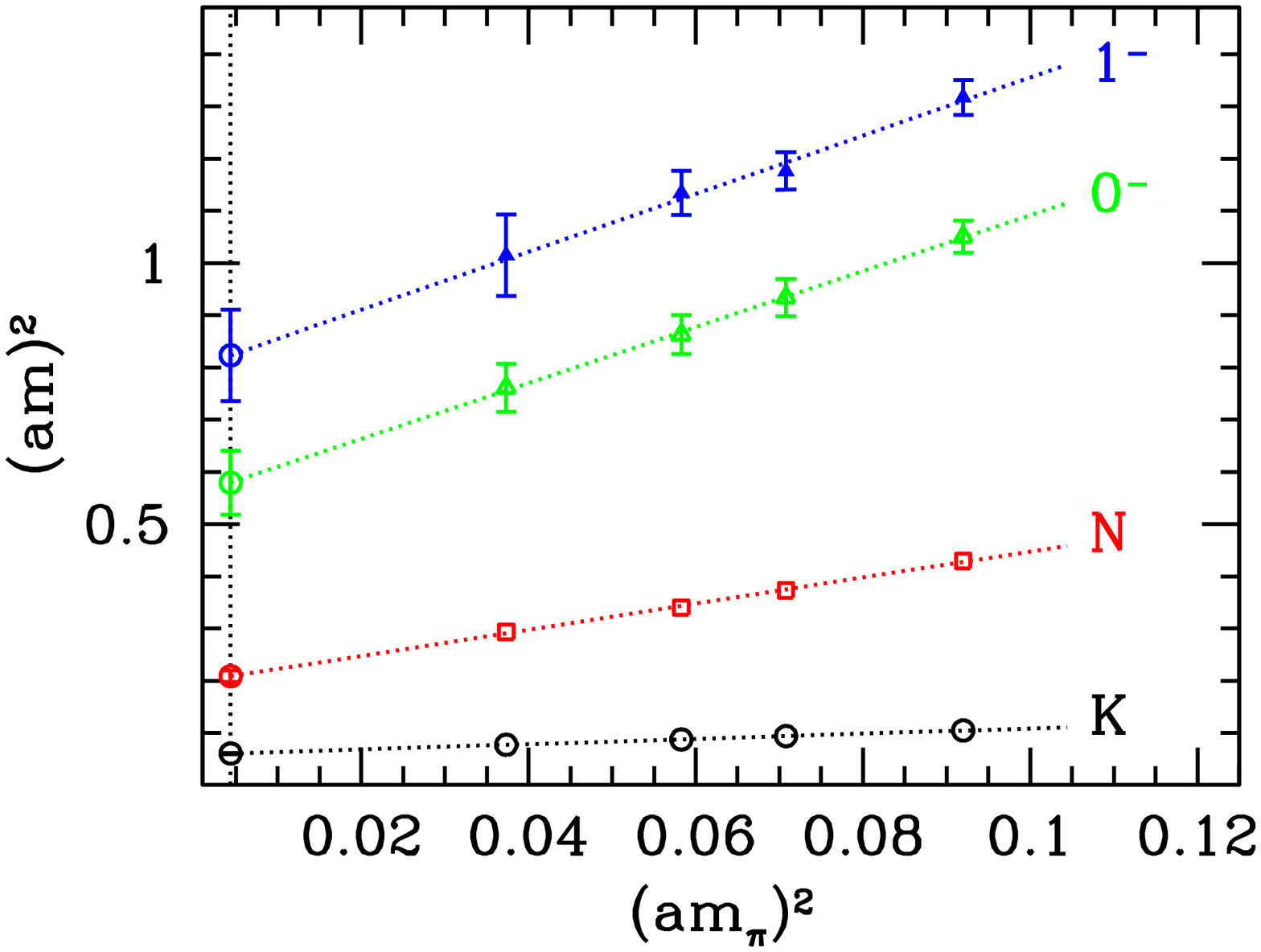}
\caption{\label{fig:chir} A typical chiral extrapolation of hadron masses
on our finest lattice at $\beta=6.0$. The particles are (from top to
bottom) the I=1,0 negative parity pentaquark states, the nucleon and
the kaon.}}
}

The chiral extrapolations for hadron masses can be performed with the 
fit functions
\be
 m_H^2 = a + b m_\pi^2 \hspace{1cm} \mbox{or} \hspace{1cm}
 m_H^2 = a + b m_\pi^2 + c m_\pi^3. 
\ee
The difference between the two extrapolations gives some information
about systematic uncertainties in the extrapolated quantities. Note,
however, that our quark masses are quite small and we have only
four different quark masses at each $\beta$. Thus with our statistics
the quadratic fit turned out to be appropriate. 
Comparing our spectrum with similar results in the literature
(c.f. \cite{Aoki:2002fd}) we concluded that 
the uncertainties due to the chiral
logarithms in the physical limit were subdominant at our present
statistics. A typical chiral extrapolation for the kaon, 
the nucleon and the negative parity $I=0,1$ pentaquark states is shown in 
Fig.\ \ref{fig:chir}. Note, that our chirally extrapolated
masses are merely illustrations. Since quenched spectroscopy is quite
reliable for mass ratios of stable particles, it is physically even more
motivated to extrapolate mass ratios instead of masses (or
equivalently one might use the squared ratios suggested
by the above mass fit formula). As
we will see the quark mass dependence of these ratios can be
weaker than that of the individual hadron masses. 

The pentaquark states have the same quark 
content as the nucleon and the kaon. Therefore, a particularly attractive 
dimensionless quantity to look at is the mass ratio of a 
pentaquark state to the nucleon plus kaon mass. We also did
chiral extrapolations of this mass ratio squared, and an example of that
for both isospins and parities is shown in Fig.\ \ref{fig:chir2}. 

\FIGURE{
\resizebox{10cm}{!}{\includegraphics{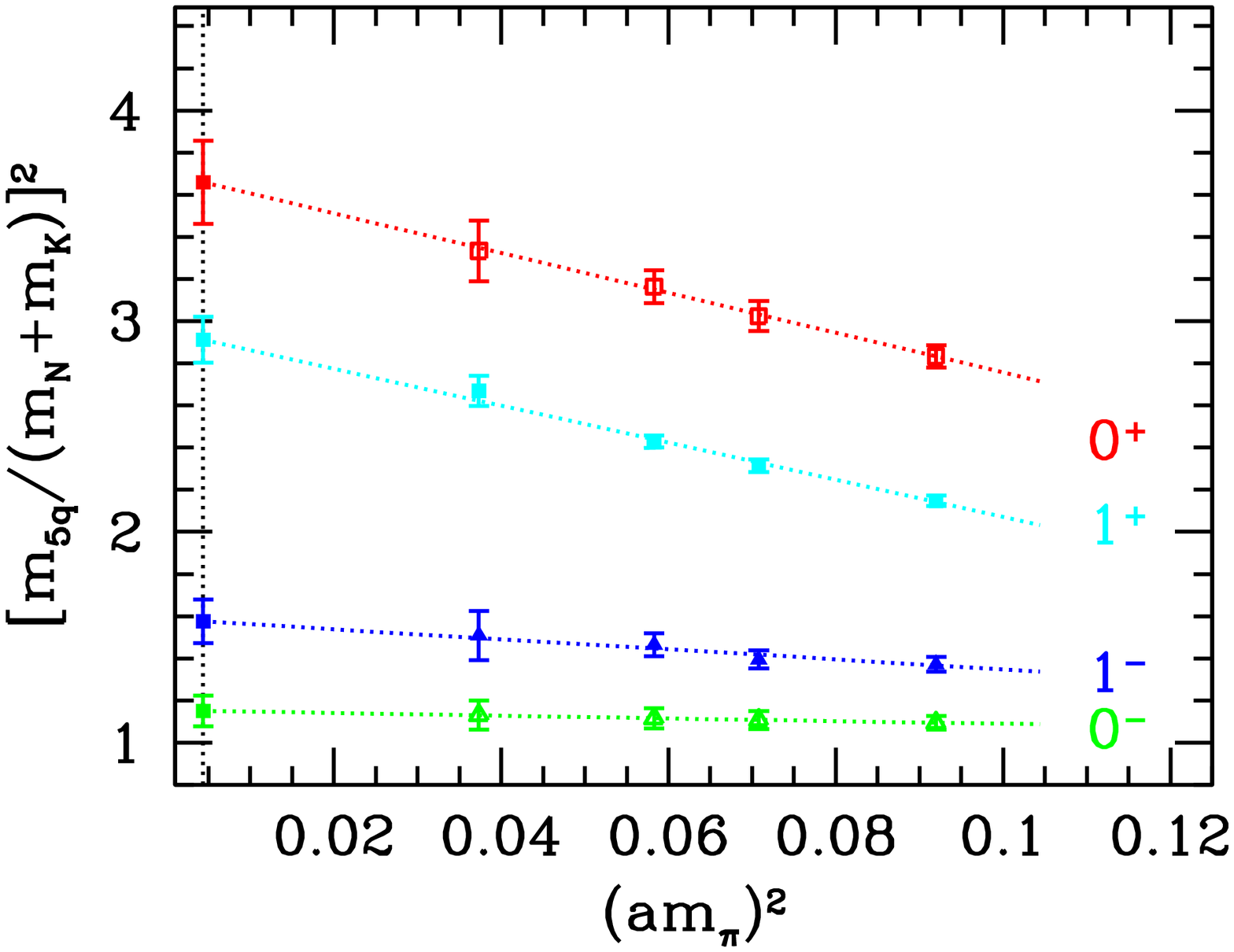}
\caption{\label{fig:chir2} Chiral extrapolation of the squared mass ratios
$m\msub{5q}^2/(m_N+m_K)^2$ on our finest lattice at $\beta=6.00$.}}
}

Finally, we performed a continuum extrapolation for the
chirally extrapolated quantities. 
In Fig.\ \ref{fig:cont} we demonstrate the continuum
extrapolation of the pentaquark to nucleon plus kaon mass ratio
for all the four pentaquark states. Since in general the Wilson
action is known to have ${\cal O}(a)$ discretization errors, we
extrapolated linearly to $a\rightarrow 0$. 
The mass ratios at hand show remarkably small scaling violations.

\FIGURE{
\resizebox{10cm}{!}{\includegraphics{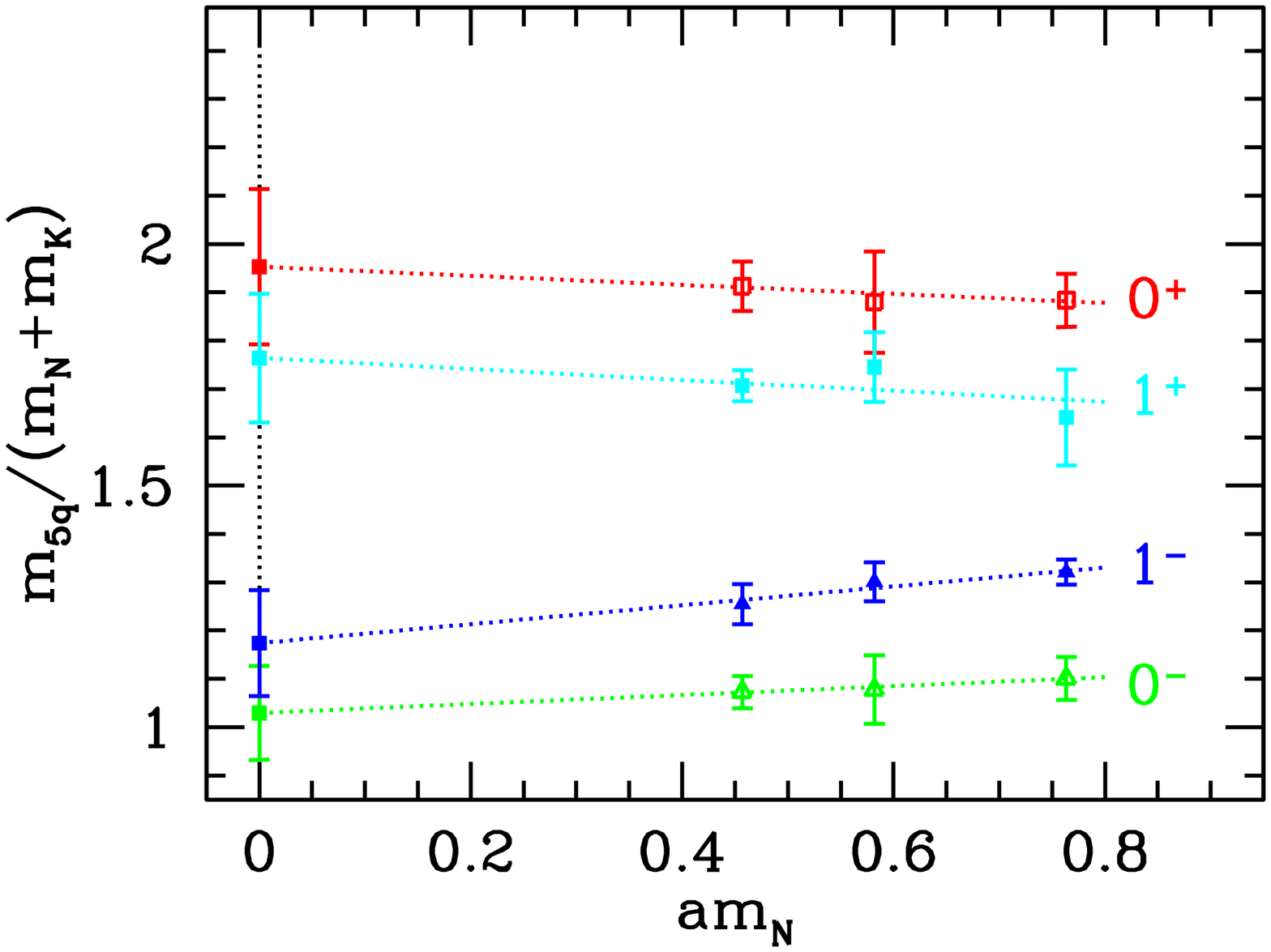}
\caption{\label{fig:cont} Continuum extrapolation of the mass ratios
$m\msub{5q}/(m_N+m_K)$ for the various pentaquark states.
The horizontal axis shows the nucleon mass in lattice units.
The lines are linear fits to the data.}}
}

\section{Results and conclusions}
     \label{se:c}

Before we present our final results we discuss how we handled 
the possible sources of
errors. In lattice QCD there are five typical sources of uncertainties:
{\it i.} statistical; {\it ii.} finite size, 
{\it iii.} quark mass, {\it iv.} finite lattice spacing and 
{\it v.} quenching errors.
We had all of these five error sources under control in our analysis. \\
{\it ad i.} 
Statistical uncertainties are well understood and they were included
by standard techniques. These techniques need uncorrelated configurations,
which we ensured by separating the analyzed 
configurations by as much as 1000 sweeps.
Statistical errors were always estimated using 
a full jackknife analysis. The statistical 
uncertainties of our pentaquark masses are typically on the few percent
level. \\
{\it ad ii.} Finite size effects in our quenched analysis turned out to
be negligible. Our spatial box size was around 2~fm. In order to look for 
finite size corrections we also performed calculations on our two 
coarsest lattices using different box sizes. With
our statistical uncertainties of order a few percent we could not
find any size dependence in the hadron masses.
\\
{\it ad iii.} Quark masses in lattice spectroscopy are usually treated
differently in the strange and in the light (up or down) sectors.  
It is straightforward to take the physical 
strange quark mass. Due to computational/technical difficulties
most of the analyses use extrapolations from light quark masses,
which are larger than the physical one. We also applied this 
method. We included all the extrapolation errors into our analysis.
\\   
{\it ad iv.} Finite lattice spacing is an inherent source of
uncertainty in any lattice analysis. We used the standard
approach. We performed our calculations at three different
lattice spacings and extrapolated to the continuum limit.
Luckily enough, the mass ratios $m_{5q}/(m_N+m_K)$
have a rather weak dependence on the lattice spacing (c.f.
Fig. \ref{fig:cont}). 
Nevertheless, the quite large, order 10\% errors on the 
finally quoted values
are mostly due to the chiral and the continuum extrapolation.
\\
{\it ad. v.} We used quenched gauge configurations, which is another 
source of systematic errors. 
Note, however, that in the case of stable hadrons, this is not
expected to be very important. Indeed, as can be seen from
Refs.\ \cite{Gattringer:2003qx} with an appropriate definition of the scale, 
the mass ratios of stable hadrons are described 
correctly by the quenched approximation on the 1-2\% level. 
We included a conservative estimate of 3\% quenching uncertainty 
in our analysis.

In addition to these five typical sources of uncertainty, there is
a somewhat special one, connected to the fact that the \Tp\ mass
is rather close to the N+K threshold. In the negative parity 
channel the S-wave scattering state might contaminate the spectrum.
Though we showed how these states can be successfully separated,
more work is needed to clarify this issue and finalise the conclusion
on the parity.

Given the fact that in the $m_{0^-}/(m_K+m_N)$ mass ratio 
we do not observe any scaling violations, we could also quote the
value of this quantity on our finest lattice, which has the smallest 
error. This would give
\be
 \frac{m_{0^-}}{m_K+m_N} = 1.073(34),
\ee
or using the physical kaon and nucleon mass
\be
 m_{0^-} = 1539 \pm 50 \mbox{MeV}.
\ee
Note, that the continuum extrapolation increases the error on 
the mass ratio, and consequently on the mass, by a factor of 3.  
   
Our final result after continuum extrapolation is shown in Fig.\
\ref{fig:result}. Of the four $I^P$ channels, the closest in mass
to the experimentally observed \Tp\ is the $0^-$ state. The $1^-$
state can be seen to be about 15\% heavier, but still within one
standard deviation of the \Tp mass. This state, however, which is
a member of an isospin triplet, 
is not the \Tp, since the latter is experimentally known
to be an isospin singlet \cite{Barth:2003es}. The parity partner 
of the $0^-$ state is almost two times heavier and lies several
standard deviations above the experimentally observed mass of the \Tp. 
This suggests that the $0^-$ state should be identified with the \Tp.

\FIGURE{
\resizebox{10cm}{!}{\includegraphics{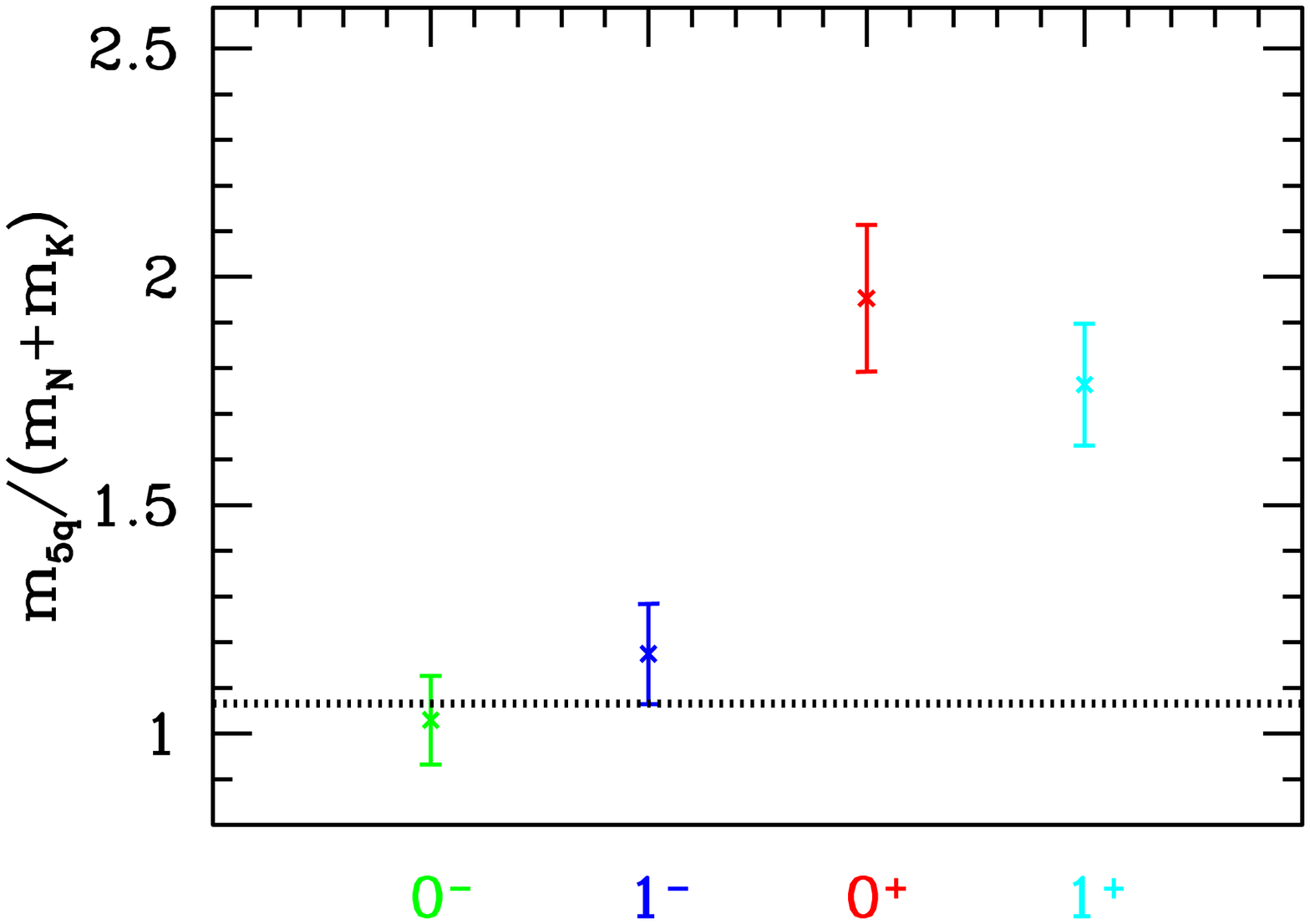}
\caption{\label{fig:result} The continuum extrapolated mass ratios
$m\msub{5q}/(m_N+m_K)$ for the lowest mass pentaquark states
in the various $I^P$ channels. The horizontal line shows the
experimentally known mass of the \Tp.}}
}

Obviously in this exploratory study we could not perform
a high statistics analysis.
Quenching effects are quite small,
nevertheless, it would be advisable to carry out the
calculations with unquenched configurations.
We did not make a systematic study of the possible
interpolating operators that
are likely to have a good overlap with the \Tp, either.
This study would be also important. Comparing different spatial and 
flavour trial wave functions is practically the only
way to obtain more information about the flavour structure of the
\Tp\ wave function and decide among the several possibilities that have
already been put forward in the recent literature (e.g.\ one could
analyze diquark-diquark-antiquark pictures \cite{Jaffe:2003sg}).
Furthermore, one expects additional exotic pentaquark hadrons with
different spin, isospin and strangeness quantum numbers.
Lattice QCD might be useful to test these light exotic
particle cascades numerically.

In conclusion, we studied spin 1/2 isoscalar and isovector candidates in both
parity channels for the recently discovered
\Tp(1540) pentaquark particle in quenched lattice QCD. 
Our analysis took into account all possible 
uncertainties.
The lowest mass that we find in the $I^P=0^-$ channel is 
in complete agreement with the experimental value of the \Tp\ mass.
On the other hand, the lowest mass state in the opposite 
parity $I^P=0^+$ channel is much higher. 
Since the $I=1$ channel has already been excluded by experiment as  
a candidate for the \Tp\ \cite{Barth:2003es}, this leaves  
the $I^P=0^-$ state to be identified with the \Tp.  
Our conclusion is consistent with that of the independent 
analysis of S. Sasaki \cite{Sasaki:2003gi},
who separated in addition also the K+N S-wave scattering state by
observing a second plateau in the effective mass.     
This successful separation is extremely important for establishing the
parity of the \Tp.
Clearly, the more sophisticated, CPU expensive, cross correlator 
technique is needed in order to unambigously separate the \Tp\ and the 
K+N S-wave and its excited states, which will finalize the conclusion 
on the mass and parity.

\acknowledgments

We would like to thank 
K.-F.~Liu, 
I.~Montvay, 
F.~Niedermayer 
and S.~Sasaki 
for useful discussions. This research was partially supported by
OTKA Hungarian Science Grants No.\ T34980, T37615, M37071, T032501.
The computations were carried out at the E\"otv\"os University 
on the 163 node PC cluster of the Department for Theoretical Physics, using
a modified version of the publicly available MILC code (see 
www.physics.indiana.edu/~sg/milc.html). T.G.K.\ also acknowledges 
support by a Bolyai Fellowship.

\end{document}